\documentclass[preprint]{aastex}

\begin{document}

\title{\bf\LARGE Spectroscopy of high proper motion  stars \\
      in the ground--based UV}


\shorttitle{Stellar spectroscopy in the ground-based UV}
\shortauthors{Klochkova et al.}

\author{V.\,Klochkova\altaffilmark{1}$^*$, T.\,Mishenina\altaffilmark{2},
 S.\,Korotin\altaffilmark{2},  V.\,Marsakov\altaffilmark{3}, \newline
 V.\,Panchuk\altaffilmark{1}, N.\,Tavolganskaya\altaffilmark{1}
 I.\,Usenko\altaffilmark{2} }

\email{$^*$  valenta@sao.ru}

\altaffiltext{1}{Special Astrophysical Observatory, Nizhnij Arkhyz, Russia}
\altaffiltext{2}{Odessa Astronomical Observatory, Odessa, Ukraine}
\altaffiltext{3}{South Federal University, Rostov-Don, Russia}

\begin{abstract}

Based on high quality spectral data (spectral resolution
R$\ge$60000) within the wavelength range of 3550--5000\,\AA{} we
determined main parameters (effective temperature, surface gravity,
microturbulent velocity, and chemical element abundances including
heavy metals from Sr to Dy) for 14 metal--deficient G--K stars with
large proper motions. The stars we studied have a wide range of
metallicity: [Fe/H]\,=\,$-0.3 \div -2.9$. Abundances of Mg, Al, Sr
and Ba were calculated with non--LTE line-formation effects
accounted for. Abundances both of the radioactive element Th and
r--process element Eu were determined using synthetic spectrum
calculations. We selected stars that belong to different galactic
populations according to the kinematical criterion and parameters
determined by us. We found that the studied stars with large proper
motions refer to different components of the Galaxy: thin, thick
disks and halo. The chemical composition of the star
BD+80$^{\circ}$\,245 located far from the galactic plane agrees with its
belonging to the accreted halo. For the giant HD\,115444 we obtained
[Fe/H]\,=\,$-2.91$, underabundance of Mn, overabundance of
heavy metals from Ba to Dy, and, especially high excess of the
r--process element Europium: [Eu/Fe]\,=\,+1.26. Contrary to its
chemical composition typical for halo stars, HD\,115444 belongs to
the disc population according to its kinematic parameters.

\end{abstract}


\newpage

\section{Introduction}

Spectral and kinematical parameters of unevolved stars that belong
to different galactic populations are the main sources of
information about chemical and kinematical evolution of the Galaxy.
We see that spectroscopy of stars of different generations in the
Galaxy is a key program for the largest telescopes in the world. During 
last decades scientists become more and more interested
in spectroscopic studies of the oldest stellar populations, since
chemical evolution of the Galaxy is imprinted in the
chemical composition of metal--poor stars of subsequent generations.
To reconstruct the chronology of the chemical and kinematic evolution of
the Galaxy we need an enormous number of high qua\-li\-ty spectral data
and accurate proper motion and parallax determination for
metal--deficient stars.

Unevolved F--G stars (dwarfs and subdwarfs) are the most effective probes
of the chemical composition and measurements of radial velocities. Their
spectra are abundant in narrow and low-blended absorptions as seen at a
high spectral resolution. In order to enlarge the sample of unblended
lines, we have to record a broad spectral region. For extremly
metal-deficient stars that have weak lines, the spectra have to be taken
in the blue and ultraviolet regions where there are more lines than in the
visible range. Considering the facts mentioned above, we performed high
resolution spectroscopy of a sample of F--G metal--poor stars at the
6--meter telescope of the Special Astrophysical Observatory of the Russian
Academy of Sciences. All our spectra were taken within a broad spectral
interval from 3550 to 5000\,\AA{}. Below we present some recent results.

\section{Spectral observations and spectra reduction}

We have selected a sample of high-velocity stars from the high proper
motion stars review~\citep{Carney}. HD/BD numbers of 14 stars are listed
in Table\,\ref{parameters}. The spectral data were taken using the
\'echelle spectrograph NES~\citep{NES} permanently mounted in a Nasmyth
focus of the 6--meter telescope of the Special Astrophysical Observatory.
NES provides a spectral resolving power of R$\ge$60000 in the spectral
range 3000--10000\,\AA{}. NES is capable to observe in the UV due to the
camera made of fused silica. Now NES works in combination with a
2048\,x\,2048~pixel CCD that has a high sensitivity in the ground-based UV
spectral range, for wavelengths longer of 3000\,\AA{}. We notice that
spectral range of the spectrograph NES is being crossed with the range
(1740--3100\,\AA{}) of the high re\-so\-lution spectrograph
UVES~\citep{UVES} which is projected for the World Space Observatory--UV
(more details see in the paper by Shustov et al.~\citeyear{Shustov}).

The 2D \'echelle--spectra were reduced (applying standard procedures
of bias subtraction, scattered light and cosmic ray trace removal,
and order extraction) using the context ECHELLE of the system MIDAS.
The context ECHELLE was modernized to process spectra obtained in
combination with an image slicer \citep{Yushkin}. The
signal-to-noise ratio for all the spectra discussed in this paper is
higher than 200. Combined with the spectral resolving power, that
allowed us not only to detect rather weak lines but also to study
their profiles.

Kinematic characteristics (proper motions and parallaxes) of the stars we
study are presented in Table\,1 in our previous
paper~\citep{SD_Vr}. In this paper, based on the same spectra taken
with the 6-meter telescope, accurate radial velocities of 15
metal--deficient stars were published~\citep{SD_Vr}. The list of
spectral lines (more than 8000) was made using the VALD
database \citep{VALD1, VALD2}. Nearly 860 unblended features from
this list were selected for Vr measurements. The standard deviation
of the measured velocity does not exceed $\sigma \le 0.9$\,km/s for
stars with the metallicity [Fe/H]$\ge -1$, and $\sigma \le 1.1$\,km/s
for stars with [Fe/H]$\le -1$.

\section{Model parameters and elemental abundances determination}

The effective temperature $T_{eff}$ was determined using the
Str\"{o}mgren $uvby\beta$ indices and calibration based on the
infrared flux method~\citep{Alonso}. Metallicity va\-lu\-es needed
for the first iteration of $T_{eff}$ determination were used from
publications, but in the following ite\-ra\-tions our spectroscopic $\rm
[Fe/H]$ values were used. In this procedure from 100 to 230 Fe {\sc
i} lines were taken into account for different stars. Additionally,
to control the $T_{eff}$ values we used the generally adopted
spectroscopic way of $T_{eff}$ determination, forcing independence
of $\lg\epsilon(Fe)$ on the lower level excitation potential.

Surface gravity was calculated using known relations:
$$\lg{\frac{g}{g_{\sun}}} = \lg{\frac{\mathcal{M}}{\mathcal{M_{\sun}}}} +
     4\lg{\frac{T_{eff}}{T_{eff_{\sun}}}}+0.4(M_{bol} - M_{bol,\sun}),$$
where $M_{bol} = V + BC + 5\lg{\pi}$ + 5, $\mathcal{M}$ -- mass of
the star,
     $M_{bol}$ is the bolometric luminosity,
     $V$ is the visual magnitude,
     $BC$ denotes bolometric correction,
     $\pi$ is the stellar parallax.

The Hipparcos parallaxes~\citep{HIP} were used for the calculations.
Bolometric corrections were calculated using a calibration formula
of Flower (\citeyear{Flower}).

The microturbulent velocity $\xi_t$ was determined for\-cing the
independence of the neutral iron abundance on the equivalent width
W$_{\lambda}$ of the line. Atmospheric parameters adopted for 14
stars are presented in Table\,\ref{parameters}. The WIDTH9 code and
Kurucz's grid of atmospheric models~\citep{KurCD13} were
used to calculate chemical abundances. To obtain the normalized
logarithmic abundance ratio of elements to iron
[Elem/Fe]\,$=\log(N_{Elem}/N_{Fe}) - \log(N_{Elem}/N_{Fe})_{\sun}$
we used solar abundances presented by Asplund et
al.~(\citeyear{Grevesse}). Spectral lines for the calculation were
selected using data from VALD \citep{VALD1, VALD2}. The atomic
constants of lines needed for abundance calculations were taken from
the same database. The list of lines was published earlier by
Klochkova et\,al.~(\citeyear{Atlas}) and it is available at {\it
http://www.chjaa.org/2006$_{-}$6$_{-}$5.html}.

\begin{figure}[hbtp]
\includegraphics[width=0.6\textwidth,bb=20 45 550 430,clip]{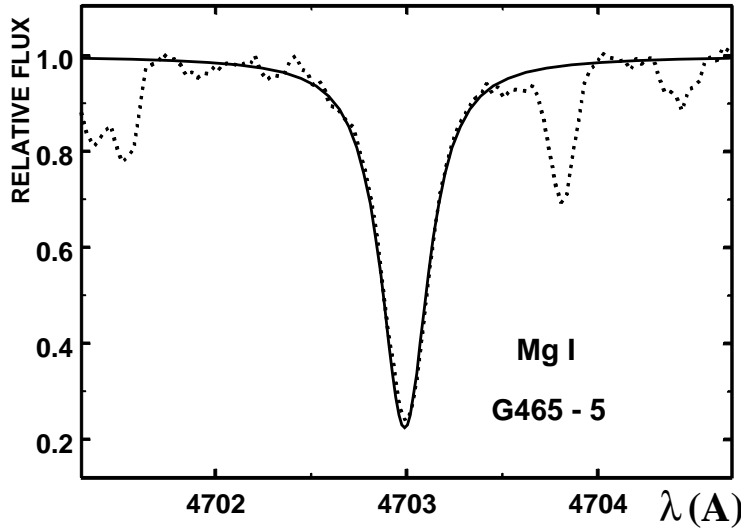}
\caption{The Mg\,{\sc I}\,$\lambda$\,4703\,\AA\ line profile (top)
        observed in the spectrum of HD\,148816 (dotted line). A synthetic
        profile is shown by a thick line}
\label{Mg4703_NLTE}
\end{figure}

The use of \'echelle spectra observed within a broad short-wavelength
spectral range allowed the model atmosphere method to be applied
to determine abundances of 20 chemical elements
including heavy me\-tals from Sr to Dy presented in Table\,\ref{Chem}.
Abundances of the iron group elements (Cr, Mn, Co, Ni, Zn), which
origin in the equilibrium Si--burning, show very small scattering
relative to the Fe abundance. Especially it concerns zinc, whose
average relative abundance is [Zn/Fe]\,=\,0.06 for stars with
[Fe/H]$ \le-1.4$.

Abundances of Mg, Al, Sr and Ba are calculated with a non--LTE
approximation (see Table\,\ref{Chem_NLTE}). The non--LTE abundance
elements was determined with a modified MULTI code~\citep{Carlsson}.
Modifications are described by Korotin et al. (\citeyear{KAL99}).
Since we use Kurucz's atmosphere models calculated with ATLAS~9
(Kurucz \citeyear{KurCD13}, \citeyear{K2005}), the necessary
background opacities for MULTI are also from ATLAS~9. In the
modified version, the mean intensities that are used to obtain the
radiative photoionization rates are calculated for a set of
frequencies at each atmospheric layer, and then they are stored in a
separate block, where they can be interpolated.

\begin{figure}[hbtp]
\includegraphics[width=0.6\textwidth,bb=20 45 550 430,clip]{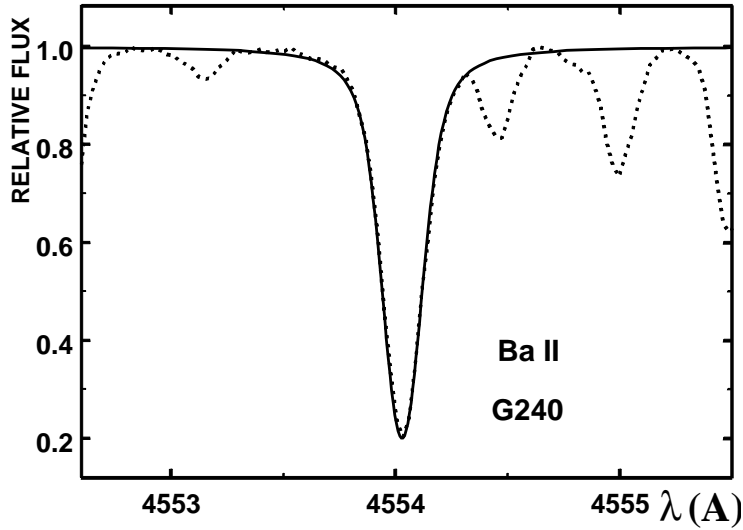}
\caption{The same as in Fig.\,1, but for the line Ba{\sc II}\,$\lambda$\,4554\,\AA{}}
\label{Ba4554_NLTE}
\end{figure}

The atomic model of magnesium used in this work is essentially the
same as that described by Mishenina et al. (\citeyear{mskk04}). This
model consists of 84 levels of Mg\,{\sc I}, 12 levels of Mg\,{\sc II}
and the ground level of Mg\,{\sc III}. Within the described system,
transitions between first 59 levels of Mg\,{\sc I} and 
ground level of Mg\,{\sc II} have been considered. The detailed
structure of the multiplets was ignored and each LS multiplet was
considered as a single term \citep[more details in][]{mskk04}.

For the non--LTE calculations we adopted an aluminium atomic model
that consists of 78 levels of Al\,{\sc I} and 13 levels of Al\,{\sc
II}. Oscillator strengths and photoionization cross sections were
taken from the TOPbase
(http://vizier.ustrasbg.fr/topbase/topbase.html). More details see
in the paper by Andrievsky et al. (\citeyear{ASKS08}).

Since the resonance Sr\,{\sc II} and Al\,{\sc I} lines in the UV band
of the spectrum are somewhat blended with other metallic lines, to
compare non--LTE line profiles with the observed spectrum one needs to
use a combination of a non--LTE and LTE synthetic spectrum. This
was done with the updated code SYNTHV \citep{TSY96} which is
designed for synthetic spectrum calculations within LTE. With this
program we calculated synthetic spectra for selected regions
comprising the Sr\,{\sc II} and Al\,{\sc I} lines of interest taking
into account all the lines from each region listed in the
VALD data-base \citep{VALD1, VALD2}. Then we included corresponding $b$-factors (factors of deviation
from LTE level populations) calculated in MULTI separately for the strontium and
aluminium lines into SYNTHV \citep{TSY96}, where they were used for calculations of the
non--LTE line source function.

Our strontium atomic model consists of 44 levels of Sr\,{\sc II} with
$n < 13$ and $l < 6$, as well as the ground level of Sr\,{\sc III}.
Since the ionization potential of the neutral strontium is only
5.7\,eV, even in atmospheres of cool stars it exists only
as of Sr\,{\sc II}. Therefore, 24 levels of Sr\,{\sc I}
were included in the model only to 
conserve the particle number. Radiative photoionization rates for the $s$, $p$ and
$d$ levels are based on the photoionization cross sections
calculated with the quantum defect method (see Andrievsky et al.
\citeyear{ASKS10} for details).

Our barium model contains 31 levels of Ba\,{\sc I}, 101 levels of
Ba\,{\sc II} with $n<50$, and the ground level of Ba\,{\sc III} ion.
We considered 91 bound--bound transitions in detail.
A cause of some uncertainty in the non--LTE analysis of the barium
spectrum is the scarce information about the photoionization
cross-sections for diffe\-rent levels. We used results obtained using the
scaled Thomas-Fermi method \citep{hof79} (for more details see
Andrievsky et al. \citeyear{ASKS09}).

To illustrate these calculations, we compare the Mg\,{\sc I}\,4702\,\AA{}
profile observed in the HD\,148816 spectrum  with a theoretical
one in Fig.\,1. A similar compa\-ri\-son for the lines
Ba\,{\sc II}\,$\lambda$\,4554\,\AA{}, Eu\,{\sc II}\,$\lambda$\,4129\,\AA{} and
Th\,{\sc II}\,$\lambda$\,4019\,\AA{} is presented in Fig.\,2--4.

Aiming to estimate cosmochronological age, the abundance ratio of
the radioactive element Th to that of the r--process element Eu is
determined for the selected stars (see Table\,\ref{Chem_Eu_Th}). Being
independent on the stellar evolution models, the cosmochronological
method of the age estimation is sensitive both to reliability of stellar
nucleosynthesis calculations and to accuracy of heavy metal
abundances derived from spectroscopic observations. We determined Eu
and Th abundances by comparison of the observed and synthetic
spectra in the region of the 4129\,\AA{} and 4019\,\AA{} lines for
Eu and Th, respectively. Synthetic spectra were calculated using a
current version of the STARSP package developed by Tsymbal
(\citeyear{TSY96}). The Th\,4019\,\AA{} line is heavily blended
(Fe\,{\sc I}, Ni\,{\sc I}, Co\,{\sc I}, V\,{\sc II}, $^{13}$CH), therefore
the error of the Th abundance determination is over 0.1 dex. Taking
into account these errors in the abundances of Eu and Th, we (as
well as other authors) obtained too large age errors, more than 4\,Gyr. For
this reason we do not present our age estimations.

\section{Discussion of results}

The stars were related to subpopulations of the Galaxy on the basis of the components of stellar spatial
velocities. Then this stratification was refined using elements of Galactic
orbits of the stars. To do this, we computed the orbital elements by simulating
30 revolutions of the star around the Galactic center, based on the
model of the Galaxy containing a disk, bulge, and extended
massive halo \citep{Allen}. The Galactocentric distance of the Sun
was assumed to be 8.5\,kpc, the rotational velocity of the Galaxy at
the solar Galactocentric distance was 220\,km/s, and the velocity of
the Sun with respect to the local standard of rest is
(U,\,V,\,W)$_{\sun} = (-11, 14, 7.5)$\,km/s \citep{Rat}. The stars
of the thin and thick disc subsystems were separated according to a
method that calculates the probability of the star membership in
each subsystem using components of their spatial velocity
\citep{Koval}. Stars of the thick disk and proper halo were
separated using total spatial velocity relative to the local
standard of rest -- V$_{\rm lsr} = 175$\,km/c \citep{Mars2005}. The
accreted halo stars (i.e. stars of extragalactic origin) were
identified using criterion -- V$_{\rm lsr} > 240$\,km/c -- these
stars have high orbital energies and most of them move on
retrograde orbits \citep{Mars2006}.

In this work, the parameters (including the metallicity) and
elemental abundances were determined for 10 stars. We also calculated the Mg, Al, Sr, Ba abundances in a non--LTE
approximation for 4 stars using parameters that were determined
earlier \cite{Atlas}. As follows from Table\,\ref{parameters}, the
stars have metallicities within a wide interval: [Fe/H]\,=\,$-0.3
\div -2.9$. The relative abundances [Elem/Fe] presented in
Tables\,\ref{Chem} and \ref{Chem_NLTE} correspond to the expected
values according to common behaviour from papers \citep{McWill,
Roederer}. This means that, for example, the $\alpha$--process
elements (Mg, Si, Ca, Ti) are overabundant for metal--poor stars at
[Fe/H]$\le -1.0$, and the value [$\alpha$/Fe] is almost constant with a
small dispersion: [$\alpha$/Fe]$=+0.3 \div 0.4$. For stars at
[Fe/H]$\ge -1.0$ abundances of these elements decrease. The star
BD+80${\rm ^o}$245 is an exception.

Our results for [Mg/Fe] are very well correspondent to the [Mg/Fe]--behaviour
derived by Gehren et al. (\citeyear{Gehren}) for an extensive stellar sample.

\begin{figure}[t!]
\includegraphics[width=0.6\textwidth,bb=60 25 720 510,clip]{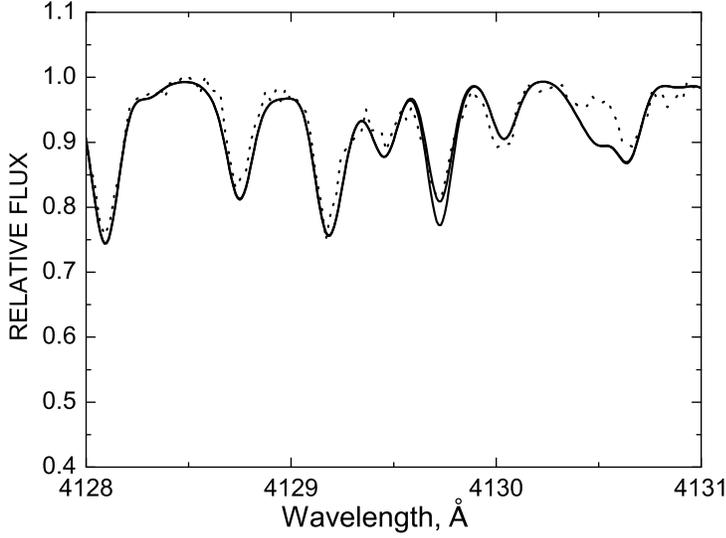}
\caption{The same as in Fig.\,1, but for the line  Eu\,{\sc II}\,$\lambda$\,4129\,\AA{}
         in the spectrum of HD\,22879}
\label{Eu_4129}
\end{figure}

\begin{figure}[t!]
\includegraphics[width=0.6\textwidth,bb=60 20 730 520,clip]{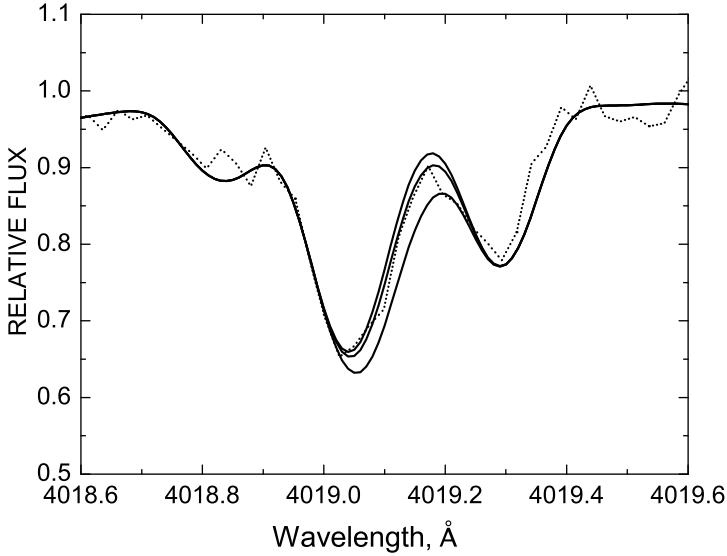}
\caption{The same as in Fig.\,1, but for the line Th\,{\sc II}\,$\lambda$\,4019\,\AA{}
        in the spectrum of HD\,5256. Synthetic spectra (from top to bottom)
	are calculated with N(Th)\,=\,$-12.5, -12.0, -11.5$}
\label{Th_G265-5}
\end{figure}

Taking into account kinematical parameters and chemical composition
of stars studied, we recognized their belonging to various galactic
populations. The most noteworthy result is a peculiar chemical
abundance pattern in the atmosphere of BD+80$^{\circ}$245 whose
metallicity is [Fe/H]$=-1.71$. The star is allocated with both
lowered (for its metallicity) abundances of Mg, Al, Ca and
remarkable deficiency of the s--procees ele\-ment Ba: [Ba/Fe]$=-1.46$.
As follows from Table\,\ref{populat}, BD+80$^{\circ}$245 is also
very distant from the Galactic plane Z$_{\rm max}=23$\,kpc.
Summarizing our results (metallicity, peculiar chemical composition,
and kinematical para\-me\-ters), we classify this star as belonging to
the accreted halo. It should be noted that this conclusion is in agreement
with earlier results of \cite{Carney97} who revealed similar
chemical peculiarities for BD+80$^{\circ}$245 and suspected that it
is an object originated not in our Galaxy. Now it is known that
the galactic halo is not a single homogeneous population, but at
least two subpopulations, outer and inner, (or initial and accreted)
\citep{Venn, Carollo}.

It should be noted that other two stars, BD+71$^{\circ}$31 and
HD\,148816 which kinematically belong to the accreted halo, do not
have chemical peculiarities similar to those obtained for
BD+80$^{\circ}$245.

There is another paradoxical object -- a giant HD\,115444 with the low  
metallicity, [Fe/H]\,=\,$-2.91$. For this star we obtained a
peculiar chemical abundance pattern which is in agreement with
published ear\-li\-er by Westin et al. (\citeyear{Westin}): 
overdeficiency of Mn and high overabundance of heavy metals from
Ba to Dy \citep{Atlas}. We noted especially high excess of the
r--process element Europium: [Eu/Fe]\,=\,+1.26. On the one hand, we
see a star with low metallicity and large overabundance of heavy
me\-tals. In total, the abundances of chemical elements are typical
for halo stars. On the other hand, kinematically HD\,115444 belongs
to the disk population. It is evident that its origin and history do
not correspond to each other.

More detailed results concerning chemical abundances, classification
by population, and elements of cosmochronology will be presented in
the forthcoming paper~\citep{SD}.

\section*{Summary}

Based on spectroscopic data (R$\ge$60000 within the wavelength range
3550--5100\,\AA{}) taken with the spectrograph NES of the 6--meter
telescope, we determined para\-me\-ters (effective temperature, surface
gravity, microturbulent velocity, metallicity, and accurate radial
velocity) of a sample of 14 metal--deficient G--K stars with large
proper motions. The use of \'echelle spectra allowed the model
atmosphere method to be applied for determination of abundances
of 20 chemical elements including heavy metals from Sr to Dy.
Abundances of Mg, Al, Sr and Ba are calculated within a non--LTE
approach.

Taking into account kinematical parameters and chemical composition of the
investigated stars, we determined galactic populations they belong to.
The chemical composition of the halo star BD+80$^{\circ}$\,245 that is
located far from the galactic plane agrees with its belonging to the
accreted halo. In addition to the overdeficiency of $\alpha$-process
elements, the star has the underabundance of the s-process element barium:
[Ba/Fe]\,=\,$-1.46$. Peculiarities of the chemical composition derived for
another metal-deficient star HD\,115444 do not correspond to its kinematic
parameters. This star also has a non-standard origin.

We conclude that the sample of high proper motion stars we studied is
inhomogeneous -- it includes objects within a wide metallicity
interval [Fe/H]\,=\,$-0.3 \div -2.9$ that belong to both disk (thin
and thick) and halo (inner and accreted).

\medskip
\section*{Acknowledgements}

V.P. and N.T. are much indebted both to the Russian Foundation for Basic
Research for financial support of this work (project 07--02--00247\,a)
and the Russian Federal program ``Observational manifestations of
evolution of chemical abundances of stars and Galaxy''. T.M. and S.K. are
grateful for the support of the Swiss national Science Fund (project
SCOPES No.\,IZ73Z0--128180/1). V.M. is grateful for the financial support
of the Ministry of education and science of the Russian Federation
(the project P.\,685).

\newpage

\begin{table}[hbtp]
\bigskip
\caption{Atmospheric parameters and metallicity [Fe/H] of program stars.
         Binary stars are marked in the last column}
\bigskip	 
\begin{tabular}{r| r|  r|  r| r| l }
\tableline
 HD, BD  &T$_{\rm eff}$& $\log$ g  &  $\xi_{\rm t}$ & [Fe/H] & \\
\tableline
    245         & 5494 & 4.20 & 0.8 &$-0.68$ & bin\\
+71${\rm ^o}$31 & 6237 & 4.24 & 1.25&$-1.83$ & bin \\
   5256         & 5212 & 3.80 & 0.5 & $-0.39$& \\
+29${\rm ^o}$366& 5609 & 4.45 & 0.5 & $-0.91$& \\
  19445         & 5890 & 4.50 & 0.7 & $-2.04$& bin?\\
  22879         & 5802 & 4.37 & 0.4 & $-0.78$& \\
 237354         & 5661 & 4.26 & 0.6 & $-0.63$& bin \\
+80${\rm ^o}$245& 5543 & 3.60 & 1.5 & $-1.71$& \\
 115444         & 4800 & 1.60 & 1.7 & $-2.91$& \\
 144061         & 5579 & 4.33 & 0.5 & $-0.31$& \\
 148816         & 5812 & 4.11 & 1.35& $-0.77$& \\
 188510         & 5410 & 5.00 & 0.6 & $-1.52$& \\
 215065         & 5567 & 4.50 & 0.5 & $-0.58$& \\
 215257         & 5900 & 4.35 & 1.0 & $-0.60$& \\
\tableline
\end{tabular}
\label{parameters}
\end{table}

\newpage

\begin{table}[hbtp]
\caption{Non-LTE abundances of Mg, Al, Sr \& Ba in the atmospheres
                 of stars studied, A(Elem). Hydrogen abundance A(H)=12.0}
\bigskip
\begin{tabular}{ l|  r|  r| r| r}
\tableline
&\multicolumn{4}{|c}{N(Elem)} \\
\cline{2-5}
Star               &Mg      &Al   & Sr  & Ba \\
\tableline
HD\,245            &7.10    &5.70 &2.07 &1.45 \\
BD+71${\rm ^o}$31  &6.05    &4.40 &1.17 &0.42 \\
HD\,5256           &7.20    &5.70 &2.32 &1.60 \\
BD+29${\rm ^o}$366 &6.88    &5.85 &1.92 &1.14 \\
HD\,19445          &6.09    &3.71 &1.02 &0.35 \\
HD22879            &7.07    &5.85 &2.07 &1.47 \\
HD\,237354         &7.00    &6.10 &2.17 &1.40 \\
BD+80${\rm ^o}$245 & 5.80   &4.60 &0.47 &$-$1.00\\
HD\,115444         &5.44    &3.45 &0.06 &$-0.06$\\
HD\,144061         &7.20    &6.00 &2.42 &1.60 \\
HD\,148816         &7.13    &6.15 &2.07 &1.27 \\
HD\,188510         &6.33    &4.03 &1.03 &0.71 \\
HD\,215065         &7.20    &6.10 &2.17 &1.40 \\
HD\,215257         &7.14    &5.23 &2.27 &1.74\\
\tableline
\end{tabular}
\label{Chem_NLTE}
\end{table}

\newpage

\begin{table*}[ht!]
\caption{Relative abundances of chemical elements [Elem/Fe]}
\bigskip
\begin{tabular}{ l|  r|  r| r| r| r| r| r| r| r| r}
\tableline
& \multicolumn{10}{c}{\small [Elem/Fe]} \\
\cline{2-11}
HD/BD&245 &+71${\rm ^o}$31 & 5256&+29${\rm ^o}$366 & 22879 &237354&+80${\rm ^o}$245& 144061& 148816  & 215065 \\
\tableline
Na{\sc I}  &  0.10 &        &  0.03  & 0.19  & 0.02  & 0.07  &       &  0.05 &  0.08   &  0.21  \\
Ca{\sc I}  &  0.09 &  0.31  &  0.02  & 0.14  & 0.10  & 0.12  &$-$0.24&$-$0.07&  0.15   &  0.11 \\
Sc{\sc II} &  0.10 &  0.23  &  0.34  & 0.23  & 0.17  & 0.28  &$-$0.31&  0.25 &  0.06   &  0.21  \\
Ti{\sc I}  &  0.09 &  0.33  &  0.08  & 0.13  & 0.23  & 0.21  &$-$0.19&  0.03 &  0.20   &  0.20  \\
Ti{\sc II} &  0.34 &  0.33  &  0.25  & 0 28  & 0 36  & 0.28  &$-$0.20&  0.18 &         &  0.25 \\
V{\sc I}   &  0.18 &  0.17  &  0.36  &       & 0.24  & 0.20  &$-$0.22&  0.26 &  0.17   &  0.32  \\
Cr{\sc I}  &  0.02 &  0.18  &$-$0.07 & 0.02  &$-$0.05& 0.03  &  0.03 &  0.12 &  0.03   &  0.10  \\
Cr{\sc II} &  0.19 &  0.08  &  0.20  & 0.10  & 0.13 & 0.22  &  0.09 &  0.31 &  0.07    &  0.20   \\
Mn{\sc I}  &$-$0.10&$-$0.18 &$-$0.03 &$-$0.2 &$-$0.25&-0.14  &$-$0.22&  0.05 &$-$0.22  &$-$0.02 \\
Fe{\sc II} &  0.05 &  0.01  &$-$0.01 &$-$0.06&$-$0.01& 0.04  &$-$0.07&  0.06 &$-$0.01  &  0.02  \\
Co{\sc I}  &  0.18 &  0.20  &  0.01  & 0.09  & 0.14  & 0.18  &$-$0.08&$-$0.04&  0.19   &  0.18  \\
Ni{\sc I}  &  0.08 &  0.15  &  0.09  & 0.02  & 0.00  & 0.09  &$-$0.05&  0.14 &  0.03   &  0.11  \\
Zn{\sc I}  &  0.39 &  0.08  &  0.47  & 0.15  & 0.16  & 0.23  &$-$0.23&  0.33 &  0.17   &  0.28  \\
Y{\sc II}  &  0.10 &  0.46  &  0.20 & 0.04  & 0.23  & 0.31  &$-$0.64&  0.23 &$-$0.07   &  0.03   \\
Zr{\sc II} &  0.09 & 0.30   &  0.22  & 0.04  & 0.21  & 0.25  &  0.24 &  0.24 &  0.04   &  0.02  \\
La{\sc II} &  0.28 &  0.44  &  0.30  & 0 22  & 0.20  & 0.13  &$-$0.41&  0.14 &         &  0.19   \\
Ce{\sc II} &  0.11 &        &  0.03  & 0.08  &$-$0.06& 0.04  &       &  0.13 &  0.02   &  0.01  \\
Nd{\sc II} &  0.23 &  0.32  &$-$0.05 & 0.17  & 0.08  & 0.19  &       &  0.22 &  0.19   &$-$0.03 \\
Sm{\sc II} &  0.26 &        &  0.15  &       & 0.25  & 0.28  &       &  0.06 &  0.10   &  0.19  \\
Gd{\sc II} &  0.41 &        &  0.09  & 0.36  & 0.07  & 0.46  &       &  0.01 &  0.11   &  0.26 \\
Dy{\sc II} &  0.46 &        &  0.46  & 0.34  & 0.34  & 0.09  &       &  0.23 & 0.06 &  0.26  \\
\tableline
\end{tabular}
\label{Chem}
\end{table*}

\newpage

\begin{table*}[t!]
\caption{Abundances $\log \epsilon$(Eu) and $\log \epsilon$(Th),
       relative abundances [Eu/H] and  [Th/H],  and ratios [Eu/Fe] and
       [Th/Fe]. Solar abundances $\log \epsilon$(Eu)\,=\,0.51 and
       $\epsilon$(Th)\,=\,0.02 are taken from \citep{Grevesse}}
\bigskip       
\begin{tabular}{ r|  r|  r|  r|  r| r|  r| r| r|  r}
\tableline
HD/BD              & 245      & 5256&+29${\rm ^o}$366& 22879 &237354 &115441 &144061 &148816 &215065 \\
\tableline
$\log\epsilon$(Eu) & 0.30     & 0.40     & 0.00    & 0.20    &0.30   &$-1.14$&0.40   &0.20   &0.30   \\
$[\rm Eu/H]$       & $-$0.21  &$-$0.11   &$-$0.51  &$-$0.31  &$-$0.21&$-1.65$&$-$0.11&$-$0.31&$-$0.21   \\
$[\rm Eu/Fe]$      & 0.47     &0.28      & 0.40    & 0.47    & 0.42  &1.26   & 0.20  &0.46   & 0.37    \\
\tableline
$\log \epsilon$(Th)& $-$0.20  &0.00      &         &$-$0.20  & 0.00  &$-1.62$&0.00   &       &  \\
$[\rm Th/H]$       & $-$0.22  &$-$0.02   &         &$-$0.22  &$-$0.02&$-1.64$&$-$0.02&       &   \\
$[\rm Th/Fe]$      & 0.46     & 0.36     &         & 0.56    & 0.61  &1.27   &0.29   &       &   \\
Th/Eu              &$-$0.50   &$-$0.40   &         &$-$0.40  &$-$0.30&$-$0.48&$-$0.40&       &   \\
\tableline
\end{tabular}
\label{Chem_Eu_Th}
\end{table*}

\clearpage
\newpage

\begin{table*}[hbtp]
\caption{Distance d, spatial velocity components U,V,W, V$_{\rm lsr}$
        and parameters of the stellar orbits calculated: e -- orbital
	eccentricity,
    Z$_{\rm max}$ -- maximum distance from the Galactic plane,
    R$_{\rm max}$ -- apogalactic radius.
    Galactic subsystems to which the stars belong is noted in the last column:
        1 -- thin disk, 2 -- thick disk, 3 -- initial halo,
    4 -- accreted halo, 1-2 -- the star rather belongs to the thin disk than
    to the thick one;
        3-2 -- the star rather belongs to the initial halo than to thick disk}
\bigskip	
\begin{tabular}{r|  r@{$\pm$}l| r|  r|  r|  r|  r| r| r| r}
\hline
HD/BD& \multicolumn{2}{c|}{d$\pm\sigma$}& U & V & W &V$_{\rm lsr}$ & e\quad& Z$_{\rm max}$ & R$_{\rm max}$ & \\
     &\multicolumn{2}{c|}{pc}&km/s &km/s &km/s &km/s  &  & kpc & kpc &       \\
\hline
     245        & 61.58& 2.7&$-$47     &$-$104 & $-$46 & 108 &0.44 & 0.6 &  9.0 & 2 \\
+71${\rm ^o}$31 &165.56&32.4&$-$171    &$-$231 & 11    & 274 &0.99 & 8.8 & 12.1 & 4 \\
    5256        & 89.37& 6.7&$-$103    &$-$77  & 44    & 125 &0.44 & 1.1 & 10.9 & 2 \\
+29${\rm ^o}$366& 56.63& 5.5&$-$63     &$-$72  & $-$52 & 94  &0.34 & 0.8 &  9.6 & 2  \\
   19445        & 38.68& 1.7&  153     &$-$120 & $-$71 &207  &0.61 & 1.5 & 11.3 & 3--2 \\
   22879        & 24.35& 0.5&$-$110    &$-$74  & $-$66 &133  &0.44 & 1.3 & 11.2 & 2 \\
  237354        & 83.96&11.4&  32      &$-$138 &$-$123 &179  &0.49 & 4.0 & 8.7  & 3--2 \\
+80${\rm ^o}$245&255.75&88.4&$-$199    &$-$367 &226    &465  &0.69 &23.0 & 28.9 & 4 \\
  115444        &281.69&97.0&  54      &$-$64  &$-$12  &84   &0.26 &0.3  & 8.9  & 1--2 \\
  144061        & 29.11& 1.0&$-$31     &$-$6   & $-$21 &26   &0.14 &0.2  & 10.3 & 1 \\
  148816        & 41.08& 1.5& 80       &$-$263 & $-$87 &280  &0.90 & 5.2 & 9.4  & 4 \\
  188510        & 39.49& 1.8&$-$167    &$-$105 & 31    &187  &0.64 &0.8  & 13.2 &3--2  \\
  215065        & 29.37& 0.5&$-$35     &$-$64  & 12    & 62  &0.27 &0.3  &9.0   & 1 \\
  215257        & 42.27& 1.7&$-$68     &19     & 37    & 78  &0.30 &1.0  &14.0  & 2 \\
\hline
\end{tabular}
\label{populat}
\end{table*}

\end{document}